\documentclass{article}

\usepackage{arxiv}

\usepackage[utf8]{inputenc} 
\usepackage[T1]{fontenc}    
\usepackage{hyperref}       
\usepackage{url}            
\usepackage{booktabs}       
\usepackage{amsfonts}       
\usepackage{nicefrac}       
\usepackage{microtype}      
\usepackage{lipsum}		    
\usepackage{graphicx}
\usepackage[english]{babel}
\usepackage{doi}
\usepackage{orcidlink} 
\usepackage{adjustbox}
\usepackage{algorithm}
\usepackage{amsmath} 
\usepackage[capitalize]{cleveref}

\title{Multi-task deep-learning for sleep event detection and stage classification}

\author{\orcidlink{0000-0003-2818-2099} Adriana Anido-Alonso \\ CITIC Research Center \\ Universidade da Coruña \\ A Coruña, 15071 \\ \texttt{adriana.anido@udc.es} \\ 
\And 
\orcidlink{0000-0001-5790-0577} Diego Alvarez-Estevez \\ CITIC Research Center \\ Universidade da Coruña \\ A Coruña, 15071 \\ \texttt{diego.alvareze@udc.es} \\ }

\date{}

\renewcommand{\headeright}{}
\renewcommand{\undertitle}{}
\renewcommand{\shorttitle}{Sleep events detection}

\begin{document}
\maketitle

\begin{abstract}
Polysomnographic sleep analysis is the standard clinical method to accurately diagnose and treat sleep disorders. It is an intricate process which involves the manual identification, classification, and location of multiple sleep event patterns. This is complex, for which identification of different types of events involves focusing on different subsets of signals, resulting on an iterative time-consuming process entailing several visual analysis passes. 
In this paper we propose a multi-task deep-learning approach for the simultaneous detection of sleep events and hypnogram construction in one single pass. Taking as reference state-of-the-art methodology for object-detection in the field of Computer Vision, we reformulate the problem for the analysis of multi-variate time sequences, and more specifically for pattern detection in the sleep analysis scenario. We investigate the performance of the resulting method in identifying different assembly combinations of EEG arousals, respiratory events ($apneas$ and $hypopneas$) and sleep stages, also considering different input signal montage configurations. 
Furthermore, we evaluate our approach using two independent datasets, assessing true-generalization effects involving local and external validation scenarios. 
Based on our results, we analyze and discuss our method's capabilities and its potential wide-range applicability across different settings and datasets.
\end{abstract}


\section{Introduction}
\label{intro}
Sleep is a fundamental physiological process for sustaining physical and mental health. For this reason, a good sleep is important to maintain a good quality of life. However, a significant proportion of the general population experiences sleep disorders that lead to worsening of the health condition \cite{medic2017short, liew2021sleep}. In order to provide accurate diagnosis and treatment of those sleep disorders, a properly sleep analysis is essential.

The standard protocol for the diagnosis of sleep disorders in Sleep Medicine involves the overnight recording of physiological activity by means of the so-called polysomnographic (PSG) test. Monitored activity includes various electroencephalographic (EEG), electrooculographic (EOG), and electromyographic (EMG) derivations, as well as recording of oro-nasal respiratory airflow and/or pressure, thoraco-abdominal movements, electrocardiogram (ECG), or blood oxygen saturation, among others \cite{aasm2023}.
Analysis of the recorded signals is afterwards performed by an expert clinician through several visual passes, looking for pattern activity of diagnostic interest, and annotating the corresponding relevant events.
More specifically, the protocol usually involves first analysis of EEG, EOG and chin EMG activity leading to the characterization of the sleep macro-structure. This task results in the so-called hypnogram, which represents the evolution of the sleep stages as a function of time. For this purpose, the PSG is segmented on a 30s epoch-by-epoch basis, each one being classified into one of the five possible sleep stages (W, N1, N2, N3, REM). Subsequently, review of the PSG is resumed back from the beginning for the detection of additional pattern activity. The process might repeat several times, entailing the identification, localization, and classification of several other events linked to specific channels, such as the presence of EEG arousals, limb movements, or different types of respiratory episodes \cite{aasm2023}. Notably, each pass involves potentially different time analysis basis (i.e. not necessarily 30s epochs), depending on the average event span and contextual information needed for the scoring of each specific target activity. 

The completion of these exhaustive manual reviews consumes a lot of time becoming unsustainable for the analyses of vast amounts of recordings. Furthermore, associated complexity leads to potential scoring inconsistencies due to clinicians fatigue and subjective interpretations \cite{younes2016staging, Cesari2021, magalang2013agreement}.

In this scenario, (semi-)automatic sleep scoring emerges as a viable solution by assisting clinicians to reduce analysis times and associated inter-rater variability \cite{alvarez-estevez_computer-assisted_2022, bakker_scoring_2022}. 
Many attempts have been made to automate several PSG analysis sub-tasks, among which sleep staging has traditionally received most of the focus \cite{schulz_rethinking_2008, phan_automatic_2022, fiorillo_automated_2019}. We also find approaches that addressed the individual detection of respiratory events \cite{yu2022sleep, zarei2022detection, liu2023detection}, EEG arousals \cite{brink2020automatic, foroughi2023deep}, or the identification of other sleep transients such as K-complexes \cite{khasawneh2023detection}. However, multi-event detection approaches are still scarce, with most of them targeting a maximum of two events simultaneously \cite{biswal2018expert, chambon2019dosed, huo2023multi, zan2023multi, liu2024msleepnet, olsen2024deep}. Consequently, concatenation of multiple independent algorithms is still needed for completion of a full PSG analysis. This leads to unpractical and sub-optimal solutions.
Moreover, the current literature often employs network architectures that rely on the concatenation of multiple independent inputs, each specialized in detecting a specific event  or depend on the creation of default templates, necessitating prior domain-specific knowledge \cite{zahid2023msed, chambon2019dosed}. Additionally, sleep staging is often overlooked as part of the analytical process, resulting in an incomplete assessment of sleep patterns. 
Furthermore, most of the related studies only consider performance evaluation on a local test partition, likely entailing dataset over-fitting risk and lack of inter-database generalization \cite{fiorillo_automated_2019, phan_automatic_2022, alvarez-estevez_challenges_2023}.

Aiming to address these problems, in this paper we introduce a novel approach for multi-event detection and hypnogram construction that takes place in a single pass. As novel contribution, our method enables the simultaneous detection of several sleep events at once, aiming to simplify the PSG analytical process and reduce the time required for patient diagnosis. Moreover, it allows for the analysis of variable input signal channels, which makes it a flexible approach to the varying availability of data.
Our method inspires from state-of-the-art methods for object detection, which are reformulated to the context of multivariate time series. 
The resulting approach is afterwards applied to sleep analysis scenario for simultaneous identification, classification, and location of three types of sleep events, regarding respiratory episodes ($apnea$ and $hypopnea$), EEG arousals, and sleep stages. Furthermore, we investigate how the proposed method performs when using different sleep event combinations as target outputs, as well as when using different input signal montages. Our approach is afterwards evaluated using a local test partition and an external, unseen, database.

In the next sections we detail each of the steps describing our method, starting from how single-shot object detection works, until all the considerations made to transfer it to our specific case. After that, we describe the experimental design and the databases used for training and validating our approach. Finally we analyze and discuss the results obtained, as well as the effects in performance when dealing with local and external testing datasets.

\section{Materials and methods}
\label{material}

\subsection{Single-shot object detection}
\label{single-shot}
\begin{figure}[t]
\centering
\includegraphics[width=0.4\textwidth]{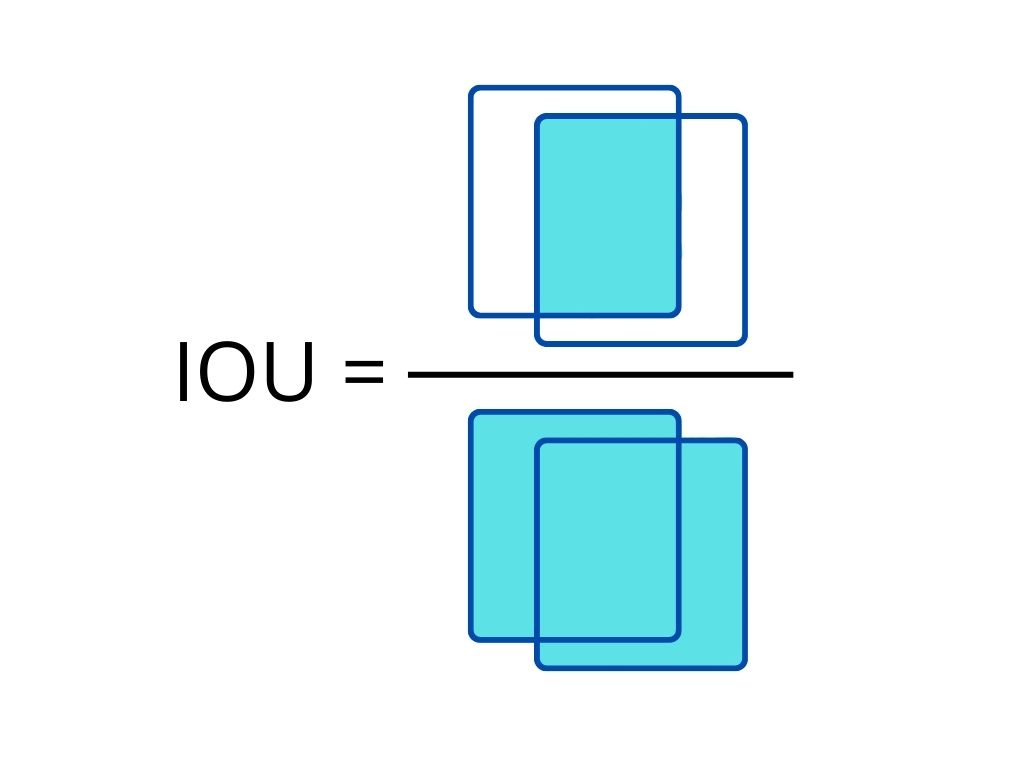}
\caption{Intersection over union (IOU).}
\label{fig:iou}
\end{figure}

In the field of Computer Vision, single-shot object detection algorithms tackle analysis of an input image for detecting different objects and their position in just one pass. You Only Look Once (YOLO), state-of-the-art algorithm in the field, belongs to such class of methods \cite{redmon_you_2016}. The YOLO algorithm processes input images using a deep convolutional neural network, dividing the image into $S \times S$ smaller cells (or grid cells), each cell being responsible for detecting possible target objects whose center falls within the margins of the cell. For this purpose, in the original version of the algorithm $B$ bounding boxes are defined within each grid cell, together with corresponding confidence scores $p$. More specifically, each grid cell $s$ is encoded as a vector of the form:
$$[p_{i}, x_{i}, y_{i}, h_{i}, w_{i}, \stackrel{\times B}{\cdots\cdots}, \overrightarrow{c_{s}}]$$

where $p_{i}$ represents the confidence score of the $i-th$ bounding box associated to the grid $s$, $x_{i}$ and $y_{i}$ its center coordinates, $h_{i}$ and $w_{i}$, respectively, the height and width of the box, and $\overrightarrow{c_{s}}$ is the associated class probability vector of size $C$, being $C$ the total number of target classes.
The confidence score $p_{i}$ reflects both how confident the model is that the corresponding box contains an object, and also how accurately does it overlap with the ground truth localization. Formally $p_{i}=Pr(Object)*IOU^{truth}_{pred}$, where IOU stands for Intersection Over the Union \cite{redmon_you_2016} (see Figure \ref{fig:iou}). 
The confidence score would be zero if no object exists in the context of the bounding boxes associated to that grill cell. Regarding the location coordinates, the original paper references $(x_{i}, y_{i})$ with respect to the corresponding grid cell, whereas $(h_{i}, w_{i})$ are taken relative to the whole image \cite{redmon_you_2016}. The associated class probabilities $c_{j}=Pr(Class_j|Object), j=1...C$, originally conditioned on the grid cell containing an object, are multiplied by the individual box confidence scores at test time: $c_{j}*p_{i}=Pr(Class_j|Object)*Pr(Object)*IOU^{truth}_{pred}=Pr(Class_j)*IOU^{truth}_{pred}$, resulting in class-specific confidence scores for each bounding box.  

In general, notice that the final length of the grid cell vector is determined at design time by fixing the number $B$ of pre-defined bounding boxes, and the number of target classes $C$, which might vary per application domain. Altogether, a final tensor of size $S \times S \times (B*5+C)$ results at the output for each input image using this setting.

Regardless, it is important to remark that subsequent versions of the algorithm \cite{terven_comprehensive_2023} introduce several modifications, some of which slightly modify the reference framework as just described. Perhaps most remarkably, in versions two and three \cite{redmon_yolo9000_2016,redmon_yolov3_2018} the class prediction mechanism is decoupled from its spatial location, allowing individual class predictions for each bounding box, instead of being linked to the grill cell. The resulting output tensor therefore would result of size $S \times S \times (B*(5+C))$. One other important modification concerns the reinterpretation of the bounding box reference coordinates in the context of anchor boxes. Using this setting, offsets with respect to pre-computed box templates are predicted instead of direct coordinates, aiming to simplify the problem and make it easier to learn for the neural network. For calculation of the $prior$ reference templates, a separated pre-calibration step, usually via k-means clustering algorithm on the training dataset, needs to be added to the pipeline.

\subsection{Multi-events detection in time sequences}
\label{multi-events}

Inspired by the aforementioned framework, we propose a method for the location and classification of event occurrences in multi-channel time series that works in one pass. For this purpose, we reformulate single-shot object detection from 2D to 1D. 

Let us consider a discrete multi-variate time series of size $D \times N$, being $D$ the number of input channels (analogously, input dimension or number of features) and $N$ the length of the series. We might split the time dimension in $S$ slices or sub-windows, in analogy to the $S \times S$ grid cells.
Following the parallelism, we are implicitly mapping the number of pixels in the image to the number $N$ of signal samples in the time series, whereas the feature dimension $D$ expands along the RGB pixel components. 
Using this setting, our proposal reinterprets "bounding boxes" as "bounding windows", effectively by dropping the $y_{i}$ and $h_{i}$ components from the positional vector. Let us reuse the notation $B$ to denote the number of (in this case) bounding windows, we characterize each window slice $s$, $s=1...S$, with the following vector: 
$$[p_{i}, x_{i}, w_{i}, \overrightarrow{c_{i}}, \stackrel{\times B}{\cdots\cdots}]$$
where $x_{i}$ and $w_{i}$ now respectively correspond to the center and the width of a 1D event spanning along the time axis.

One should notice, following this characterization, that in our proposal we are coupling the class prediction mechanism to the bounding window through $\overrightarrow{c_{i}}$, rather than to the whole window slice $s$. Still, the slice window remains responsible for detecting events whose center falls within its corresponding onset and offset intervals. The final output tensor shape for each input $D \times N$ time series following the proposed approach hence results $S \times (B*(3+C))$.

At inference time, one relevant consideration regards the differentiation of individual events of the same class, either within, or spanning across several window slices. In our proposed framework we reinterpret IOU along the temporal axis only, i.e. in terms of bounding window overlapping, subsequently applying non-maximal suppression accordingly, using pre-defined overlapping threshold $\lambda$.

\subsection{Case of study: sleep medicine}
\label{case_of_study}

Here we further specify the general-purpose multi-variate analysis framework described in the previous section and adapt it to PSG analysis in sleep medicine. More specifically, in this work we aim for detection of sleep stages, EEG arousals and respiratory events in one single pass.

Because one single PSG involves a minimum of 6-8 hours of recorded sleep data at sampling rates ranging 100-500 Hz, tackling the whole PSG as one single input becomes computationally intractable. Rather, much like a video is composed as sequence of images, we might envision PSG as a sequence of PSG sub-intervals, each one of size $D \times N$ which constitute the individual inputs to our method. In this case $D$ represents the number of signal derivations and $N$ the length of the PSG sub-interval. 
In this work we fix $N$ to be 30s and experiment with different number of input channels $D$. This is further discussed in the next section. 

For this preliminary work we also set $S$ (i.e. the number of sub-window slices) to be 1, and define a total of $B=3$ bounding windows.
Each window is responsible for one of the targeted detection outcomes (i.e. sleep stages, EEG arousals, and respiratory events). The reason for this linking is justified by the differences in the shape of the corresponding bounding windows. Let us denote by $s$, $a$, and $r$ subscripts to refer to each target outcome respectively. In the case of sleep stages we drop the $p_{s}$, $x_{s}$, and $w_{s}$ coordinates, because it follows from the reference clinical scoring procedures \cite{aasm2023}, that on a 30s epoch window one of the five possible sleep stages (W, N1, N2, N3, REM) must exist, spanning the duration of the whole window. Consequently in this case we are only left with $|\overrightarrow{c_{s}}|=5$. For EEG arousals, we are only concerned with the presence or absence of event. Thus, $|\overrightarrow{c_{a}}|=1$, which effectively becomes scalar $c_{a}=\{0,1\}$, and we can merge it with corresponding $p_{a}$. Finally, for respiratory events we use the standard $3+C$ bounding window, i.e. $[p_{r}, x_{r}, w_{r}, \overrightarrow{c_{r}}]$. More specifically, in this work we consider two different types of respiratory events, namely $apneas$ and $hypopneas$, and therefore $|\overrightarrow{c_{r}}|=2$.

Put all together, the final output tensor becomes a vector of length 13 with the following form:
\begin{align*} 
[c_{s}^{W}, c_{s}^{N1}, c_{s}^{N2}, c_{s}^{N3}, c_{s}^{R}, c_{a}, x_{a}, w_{a}, p_{r}, c_{r}^{a}, c_{r}^{h}, x_{r}, w_{r}].
\end{align*}

During our experiments we will also consider partial versions of this vector as a consequence of taking into account different combinations of target outcomes. The details are given in the corresponding section.

\subsection{Deep learning model structure}
\label{model}

A convolutional (CNN) long short-term memory (LSTM) deep-learning model is used adapted from a previous work that focused exclusively on classification of sleep stages \cite{alvarez-estevez_inter-database_2021}. Briefly, the model is divided into three main modules: a pre-processing module, a convolutional module and a multi-output LSTM module.

The first module receives as input a tensor of size $D \times M$, with $D$ representing signals of length $M$. More specifically, $M=N\pm\delta$, with $\delta$ corresponding to a configurable parameter that expands the input on both sizes around our target analysis window $N$. This is aimed for providing additional contextual information to the neural network, thus resulting in final input tensors of size $D\times(2*\delta + N)$. All signals are resampled at 100Hz for homogenization purposes and standardized using a Gaussian normalization. No artifact removal nor additional filtering is applied in this work. 

Input sequences are then split into $P$ parts, time distributed to the second module composed of three consecutive convolution blocks involving a convolutional step with ReLu activation and average pooling. We use a kernel size of 100, consistently maintained across all three convolutional blocks, but the number of filters is doubled on each block from 8 to 32. We deliberately avoided the use of batch normalization layers at the end of the block, as it proved problematic for some particular generalization settings \cite{anido2023decentralized}. Final features are extracted of the final convolutional block, flattening the resulting tensor followed by a dense layer with output length 50 and 0.5 rate drop-out. 
Finally, features resulting from each time distributed block are fed into a LSTM block for learning time dependencies followed by dense connectivity to the final output layer. This final layer uses sigmoid and softmax activations, respectively for binary and multi-label classification neurons (i.e. $c_{i}$ and $p_{i}$ components of the output vector), while linear activation is used for location coordinates (that is, $x_{i}$ and $w_{i}$).  In total, ten experiments are scheduled that result from mashup of the described configurations. 
The architecture has been implemented in Python and Tensorflow (versions 3.11.4 and 2.12.0, respectively).

\subsection{Databases}
\label{databases}
To conduct the experiment, two independent and heterogeneous databases where used. 

\textit{Sleep Heart Health Study database (SHHS)}. SHHS contains data from a multi-center cohort study implemented by National Heart, Lung and Blood Institute, with the objective of analyzing cardiac consequences of sleep-disordered breathing \cite{quan_shhs_1997}. It contains polysomnographic recordings from 6.441 adults tracked until 2010, divided in two different visits SHHS-1 and SHHS-2. The database is available online upon request at the National Sleep Research Resource (NSRR) \cite{zhang2018national}. For this study, a random subset of 100 recordings from the SHHS-2 cohort was selected to train and locally validate our proposed method. This selection is motivated by the need to establish a baseline for comparison in the detection of sleep stages, in line with previous research \cite{alvarez-estevez_inter-database_2021, anido2023decentralized}.

\textit{Haaglanden Medisch Centrum Inter-Scorer Analysis database (HMC-ISA)}.
This dataset includes 5 PSG recordings that were part of a previous study aimed at evaluating inter-rater variability in the scoring of various PSG event annotations. More specifically, recordings selected for this study correspond to the 5 PSG subgroup originally targeting the assessment of variability associated to the leg movement's scoring subtask. For this task pre-filled annotations related to scoring of hypnogram, EEG arousals, and respiratory event's activity were provided as baseline information to help the scorers' contextual interpretation. These data are used in this study as the reference target output for assessing the external dataset generalization of our method. Further details about the study and related data can be found in \cite{alvarez-estevez_computer-assisted_2022}.

\section{Experimental design}
\label{experiments}

We consider three different subset combinations of input channels to observe their influence at predicting the different targeted event assemblies. 
More specifically, the combinations (2$\times$EEG, EOG, EMG), (2$\times$EEG, EOG, EMG, saturation, airflow) and (2$\times$EEG, EOG, EMG, saturation, airflow, abdominal movements, thoracic movements) are used, resulting in corresponding input feature dimensionalities $D = 4$, $D = 6$, and $D = 8$. Different combinations of targeted assemblies are also examined, involving the detection of sleep stages, EEG arousals and respiratory events. In particular, sleep stages and EEG arousals ($s + a$), sleep stages and respiratory events ($s + r$), and all together ($s + a + r$). In addition, we consider the solely detection of sleep stages ($s$) for $D = 4$ as baseline reference. In total, ten experiments are scheduled that result from mashup of the described configurations. 

For all the experiments we fix the reference analysis window $N = 30s$, intending to match the time basis used by clinicians to annotate sleep stages. Inputs to the neural network are then constructed by expanding the reference window 60s on both sides ($\delta = 60$). This is done with the objective to provide the neural network with additional temporal context for sequential analysis, mimicking clinical procedures \cite{aasm2023}. For this work, we also considered five splits ($P = 5$) for time distribution to the LSTM module of the proposed deep learning architecture, as described in the corresponding section. 
Also, as previously described, no sub-windowing is considered for this trial ($S = 1$), defining up to three bounding windows ($B$), depending on the considered output assembly configuration, each one being responsible for detection of one specific event type. 

All the experiments use the set of databases described in section \ref{databases}, with SHHS used for training and local performance evaluation, and HMC-ISA for assessment of inter-database generalization capabilities. In this regard, we use an epoch-wise 80-20\% split partition scheme using the SHHS database, where 20\% is set aside as local test dataset (TS), whereas the remaining 80\% is further split into another 80-20\% comprising training (TR) and validation (VAL) sets. In contrast, the complete set of HMC-ISA (FULL) is used for external generalization performance evaluation.

During training, input samples are processed in batches of size 100. We configure a large-enough number of maximum iterations (max. epochs = 100) and enable early stopping based on loss performance evaluated in VAL. We set a patience of 5 in this regard.

We use a three-component loss function involving binary and categorical cross-entropy, respectively for binary and multi-class outputs, and Mean Squared Error (MSE) for location coordinates. The use of three separated losses was motivated by experimental results in this work showing an increase in performance compared to the use of a single-component loss function. This effect will be further analyzed in section \ref{discussion}. We do not utilize any weighting mechanism to moderate losses effects during the learning process, both tasks are aggregated and contribute equally. This is motivated as we would like to treat each task with equal importance in the learning process.
To optimize our objective function we use stochastic gradient descent with a constant learning rate (lr = 0.001) and momentum (p = 0.9). 

Outcomes for each experiment are assessed both, on local and external datasets, using Cohen’s Kappa ($\kappa$), F1-Score and Mean Absolute Error (MAE). We ought Kappa as the reference performance metric for evaluation of event detection due to the multi-class imbalanced nature of the different sleep event categories, for which this metric corrects for agreement due to chance. Furthermore, Kappa is a widespread metric for examining inter-rater agreement in sleep scoring tasks across the reference literature. We also select F1-Score that combines both classification precision and recall in one metric. Finally, MAE is deemed adequate for interpretation of regression sub-task regarding temporal event localization, and better global overview of model's performance across different experiments (classification and regression) by analyzing absolute error between the predicted outputs versus true labels.  

Source code for experimental reproducibility of this study is available at GitHub (\href{https://github.com/adrania/sleep-events-detection.git}{https://github.com/adrania/sleep-events-detection.git}) and Zenodo \cite{anido_alonso_2024_14505513}.

\subsection{Datasets generation}
\label{datasets}

We create ten datasets from SHHS and HMC-ISA according to the experimental design. 
For this purpose, each PSG is split into $D \times N$ sub-intervals for which desired output is constructed based on the available set of labels resulting from clinical expert's event annotations, and the specific targeted output assembly configuration, i.e. ($s$), ($s$ + $a$), ($s$ + $r$) or ($s$ + $a$ + $r$).

Because sub-intervals and annotations are related to a continuous PSG record, it is essential to maintain the intrinsic temporal correspondence between both sequences. Therefore, we refer the margins of each sub-interval to their precise location in the PSG. Accordingly, we associate each annotation to a single sub-interval by calculating its centroid $z = \frac{i + t}{2}$, where $i$ represents the corresponding event annotation onset with respect to the entire PSG, and $t$ its duration. 
Consequently, if an annotation spans several sub-intervals, only the one that contains its centroid will be responsible for its detection.
Therewith, the desired output vector is filled with the corresponding expert's scores regarding the event's presence or absence ($p_{i}$), its corresponding class ($\overrightarrow{c_{i}}$), and its location ($x_{i}$ and $w_{i}$), hereafter referred as $B_{i}$. Note that, if there is no annotation in the current sub-interval, $p_{i}$, $\overrightarrow{c_{i}}$ and $B_{i}$ are assigned with 0. 
We only consider the occurrence of one annotation per class on each sub-interval, therefore, if two annotations of the same class fall in the current sub-interval, the one with longer duration is prioritized.
Note as well that centroid calculation is not applicable to sleep stages, since assuming one stage per sub-interval, expert annotations perfectly correlate with each $D$ x $N$ sub-interval.
In contrast, input patterns are extended $\delta$ on each side and normalized resulting in zero-mean and unit standard deviation per channel. Similarly, bounding window coordinates are normalized with respect to the reference sub-interval limits.

At inference time, sigmoid and softmax activations are turned into categorical values following one-hot encoding (0 or 1). For sigmoid activations (i.e. $p_{i}$) we use a 0.5 threshold to categorize the presence or absence of the current event. Thereby, if the activation value exceeds that threshold it is considered as event presence ($p_{i}$ = 1) and vice versa as event absence ($p_{i}$ = 0). In the latter case, its corresponding $\overrightarrow{c_{i}}$ and $B_{i}$ are turned to 0.
Similarly, we extract the corresponding event class by selecting the position of $\overrightarrow{c_{i}}$ with higher softmax activation, which is set to 1, while the remaining positions are set to 0.
We also apply additional post-processing on the inferenced events following AASM specifications \cite{aasm2023}. For instance, inferenced EEG arousals whose duration last less than 3 seconds are discarded. Similarly, events of the same class that overlap are treated as one single annotation.

\begin{table*}[!]
\caption{Comparison between the use of a single loss function (MSE) and the use of dedicated three component loss (MSE + binary cross-entropy + categorical cross-entropy) for the simultaneous detection of sleep stages, EEG arousals and respiratory events. Results correspond to simulation involving $D = 8$ input channels on the local TS SHHS partition. TS = testing Set; TR = training; c = classification; bw = bounding window; MAE = Mean Absolute Error. Notice that because sleep stages does not involve event localization, corresponding MAE refers to the classification task only (MAE c). For respiratory and EEG arousal events MAE refers to the classification (MAE c) and detection (MAE bw) tasks.}
\begin{center}
\begin{adjustbox}{width=\textwidth}
\begin{tabular}{@{}ccccccccccccccc@{}}
 &  & \multicolumn{3}{c}{\textbf{SLEEP STAGES (SHHS TS)}} & \multicolumn{4}{c}{\textbf{EEG AROUSALS (SHHS TS)}} & \multicolumn{4}{c}{\textbf{RESPIRATORY EVENTS (SHHS TS)}} & \textbf{} &  \\ \midrule
\textbf{Model} & \textbf{Loss} & \textbf{Kappa} & \textbf{F1 Score} & \textbf{MAE c} & \textbf{Kappa} & \textbf{F1 Score} & \textbf{MAE c} & \textbf{MAE bw} & \textbf{Kappa} & \textbf{F1 Score} & \textbf{MAE c} & \textbf{MAE bw} & \textbf{global MAE} & \textbf{TR epochs} \\ \midrule
M10.1 & \multicolumn{1}{c|}{single} & 0,78 & 0,69 & \multicolumn{1}{c|}{0,06} & 0,52 & 0,76 & 0,11 & \multicolumn{1}{c|}{0,07} & 0,61 & 0,73 & 0,11 & \multicolumn{1}{c|}{0,11} & 0,08 & 55 \\
M10.2 & \multicolumn{1}{c|}{multiple} & 0,80 & 0,77 & \multicolumn{1}{c|}{0,04} & 0,58 & 0,79 & 0,08 & \multicolumn{1}{c|}{0,06} & 0,63 & 0,75 & 0,08 & \multicolumn{1}{c|}{0,09} & 0,06 & 31 \\ \bottomrule
\end{tabular}
\end{adjustbox}
\end{center}
\label{table_loss}
\end{table*}

\begin{table*}[!]
\caption{Models performance on predicting local SHHS test partition (TS). Each model predicts different combinations of sleep events (E) involving sleep stages (s), EEG arousals (a) and respiratory events (r), using different amount of input channels (D). Metrics are shown by event type. c = classification; bw = bounding window; MAE = Mean Absolute Error. Notice that because sleep stages does not involve event localization, corresponding MAE refers to the classification task only (MAE c). For respiratory and EEG arousal events MAE refers to the classification (MAE c) and detection (MAE bw) tasks.}
\begin{center}
\begin{adjustbox}{width=\textwidth}
\begin{tabular}{@{}ccccccccccccccc@{}}
 &  & \textbf{} & \multicolumn{3}{c}{\textbf{SLEEP STAGES (SHHS TS)}} & \multicolumn{4}{c}{\textbf{EEG AROUSALS (SHHS TS)}} & \multicolumn{4}{c}{\textbf{RESPIRATORY EVENTS (SHHS TS)}} & \textbf{} \\ \midrule
\textbf{Model} & \textbf{D} & \textbf{E} & \textbf{Kappa} & \multicolumn{1}{l}{\textbf{F1 Score}} & \textbf{MAE c} & \textbf{Kappa} & \multicolumn{1}{l}{\textbf{F1 Score}} & \textbf{MAE c} & \textbf{MAE bw} & \textbf{Kappa} & \multicolumn{1}{l}{\textbf{F1 Score}} & \textbf{MAE c} & \textbf{MAE bw} & \textbf{global MAE} \\ \midrule
M1 & 4 & \multicolumn{1}{c|}{s} & 0,79 & 0,74 & \multicolumn{1}{c|}{0,06} & - & - & - & \multicolumn{1}{c|}{-} & - & - & - & \multicolumn{1}{c|}{-} & 0,06 \\
M2 & 4 & \multicolumn{1}{c|}{s + a} & 0,82 & 0,77 & \multicolumn{1}{c|}{0,05} & 0,70 & 0,85 & 0,08 & \multicolumn{1}{c|}{0,06} & - & - & - & \multicolumn{1}{c|}{-} & 0,05 \\
M3 & 4 & \multicolumn{1}{c|}{s + r} & 0,83 & 0,79 & \multicolumn{1}{c|}{0,05} & - & - & - & \multicolumn{1}{c|}{-} & 0,48 & 0,63 & 0,15 & \multicolumn{1}{c|}{0,14} & 0,10 \\
M4 & 4 & \multicolumn{1}{c|}{s + a + r} & 0,82 & 0,78 & \multicolumn{1}{c|}{0,05} & 0,68 & 0,84 & 0,08 & \multicolumn{1}{c|}{0,06} & 0,47 & 0,62 & 0,15 & \multicolumn{1}{c|}{0,13} & 0,09 \\
M5 & 6 & \multicolumn{1}{c|}{s + a} & 0,82 & 0,79 & \multicolumn{1}{c|}{0,05} & 0,62 & 0,81 & 0,10 & \multicolumn{1}{c|}{0,06} & - & - & - & \multicolumn{1}{c|}{-} & 0,06 \\
M6 & 6 & \multicolumn{1}{c|}{s + r} & 0,81 & 0,77 & \multicolumn{1}{c|}{0,05} & - & - & - & \multicolumn{1}{c|}{-} & 0,60 & 0,73 & 0,12 & \multicolumn{1}{c|}{0,12} & 0,08 \\
M7 & 6 & \multicolumn{1}{c|}{s + a + r} & 0,83 & 0,79 & \multicolumn{1}{c|}{0,05} & 0,61 & 0,80 & 0,09 & \multicolumn{1}{c|}{0,06} & 0,61 & 0,73 & 0,11 & \multicolumn{1}{c|}{0,11} & 0,08 \\
M8 & 8 & \multicolumn{1}{c|}{s + a} & 0,79 & 0,75 & \multicolumn{1}{c|}{0,06} & 0,47 & 0,73 & 0,12 & \multicolumn{1}{c|}{0,07} & - & - & - & \multicolumn{1}{c|}{-} & 0,07 \\
M9 & 8 & \multicolumn{1}{c|}{s + r} & 0,82 & 0,78 & \multicolumn{1}{c|}{0,04} & - & - & - & \multicolumn{1}{c|}{-} & 0,65 & 0,75 & 0,08 & \multicolumn{1}{c|}{0,09} & 0,06 \\
M10 & 8 & \multicolumn{1}{c|}{s + a + r} & 0,82 & 0,79 & \multicolumn{1}{c|}{0,05} & 0,60 & 0,80 & 0,10 & \multicolumn{1}{c|}{0,07} & 0,65 & 0,77 & 0,10 & \multicolumn{1}{c|}{0,10} & 0,08 \\ \midrule
 &  & Average & 0,82 & 0,78 & 0,05 & 0,61 & 0,81 & 0,10 & 0,06 & 0,58 & 0,70 & 0,12 & 0,11 & 0,07
\end{tabular}
\end{adjustbox}
\end{center}
\label{local_performance}
\end{table*}

\begin{table*}[!]
\caption{Performance of each pre-trained model when predicting the entire (FULL) set of the external HMC-ISA database. Each model predicts different combinations of sleep events (E) involving sleep stages (s), EEG arousals (a) and respiratory events (r), using different amount of input channels (D). Metrics are shown by event type. c = classification; bw = bounding window; MAE = Mean Absolute Error. Notice that because sleep stages does not involve event localization, corresponding MAE refers to the classification task only (MAE c). For respiratory and EEG arousal events MAE refers to the classification (MAE c) and detection (MAE bw) tasks.}
\begin{center}
\begin{adjustbox}{width=\textwidth}
\begin{tabular}{@{}ccccccccccccccc@{}}
 &  & \textbf{} & \multicolumn{3}{c}{\textbf{SLEEP STAGES (HMC-ISA FULL)}} & \multicolumn{4}{c}{\textbf{EEG AROUSALS (HMC-ISA FULL)}} & \multicolumn{4}{c}{\textbf{RESPIRATORY EVENTS (HMC-ISA FULL)}} & \textbf{} \\ \midrule
\textbf{Model} & \textbf{D} & \textbf{E} & \textbf{Kappa} & \multicolumn{1}{l}{\textbf{F1 Score}} & \textbf{MAE c} & \textbf{Kappa} & \multicolumn{1}{l}{\textbf{F1 Score}} & \textbf{MAE c} & \textbf{MAE bw} & \textbf{Kappa} & \multicolumn{1}{l}{\textbf{F1 Score}} & \textbf{MAE c} & \textbf{MAE bw} & \textbf{global MAE} \\ \midrule
M1 & 4 & \multicolumn{1}{c|}{s} & 0,48 & 0,53 & \multicolumn{1}{c|}{0,16} & - & - & - & \multicolumn{1}{c|}{-} & - & - & - & \multicolumn{1}{c|}{-} & 0,16 \\
M2 & 4 & \multicolumn{1}{c|}{s + a} & 0,56 & 0,60 & \multicolumn{1}{c|}{0,13} & 0,26 & 0,63 & 0,12 & \multicolumn{1}{c|}{0,06} & - & - & - & \multicolumn{1}{c|}{-} & 0,11 \\
M3 & 4 & \multicolumn{1}{c|}{s + r} & 0,47 & 0,52 & \multicolumn{1}{c|}{0,16} & - & - & - & \multicolumn{1}{c|}{-} & 0,05 & 0,34 & 0,19 & \multicolumn{1}{c|}{0,11} & 0,16 \\
M4 & 4 & \multicolumn{1}{c|}{s + a + r} & 0,56 & 0,59 & \multicolumn{1}{c|}{0,14} & 0,23 & 0,61 & 0,13 & \multicolumn{1}{c|}{0,06} & 0,05 & 0,32 & 0,19 & \multicolumn{1}{c|}{0,11} & 0,13 \\
M5 & 6 & \multicolumn{1}{c|}{s + a} & 0,43 & 0,50 & \multicolumn{1}{c|}{0,18} & 0,22 & 0,61 & 0,16 & \multicolumn{1}{c|}{0,08} & - & - & - & \multicolumn{1}{c|}{-} & 0,15 \\
M6 & 6 & \multicolumn{1}{c|}{s + r} & 0,40 & 0,48 & \multicolumn{1}{c|}{0,19} & - & - & - & \multicolumn{1}{c|}{-} & 0,34 & 0,52 & 0,07 & \multicolumn{1}{c|}{0,04} & 0,12 \\
M7 & 6 & \multicolumn{1}{c|}{s + a + r} & 0,54 & 0,58 & \multicolumn{1}{c|}{0,14} & 0,24 & 0,62 & 0,11 & \multicolumn{1}{c|}{0,06} & 0,30 & 0,50 & 0,09 & \multicolumn{1}{c|}{0,05} & 0,10 \\
M8 & 8 & \multicolumn{1}{c|}{s + a} & 0,35 & 0,44 & \multicolumn{1}{c|}{0,21} & 0,20 & 0,60 & 0,11 & \multicolumn{1}{c|}{0,06} & - & - & - & \multicolumn{1}{c|}{-} & 0,16 \\
M9 & 8 & \multicolumn{1}{c|}{s + r} & 0,38 & 0,45 & \multicolumn{1}{c|}{0,20} & - & - & - & \multicolumn{1}{c|}{-} & 0,36 & 0,53 & 0,07 & \multicolumn{1}{c|}{0,04} & 0,13 \\
M10 & 8 & \multicolumn{1}{c|}{s + a + r} & 0,29 & 0,38 & \multicolumn{1}{c|}{0,23} & 0,21 & 0,60 & 0,12 & \multicolumn{1}{c|}{0,06} & 0,13 & 0,42 & 0,17 & \multicolumn{1}{c|}{0,09} & 0,16 \\ \midrule
 &  & Average & 0,45 & 0,51 & 0,17 & 0,23 & 0,61 & 0,13 & 0,06 & 0,21 & 0,44 & 0,13 & 0,08 & 0,14
\end{tabular}
\end{adjustbox}
\end{center}
\label{external_performance}
\end{table*}

\section{Results}

Results of the scheduled experiments are detailed in \crefrange{table_loss}{external_performance}, and in Figures \ref{fig:local_mae} and \ref{fig:external_mae}.

Table \ref{table_loss} presents the performance comparison between single- and three-component loss function for detecting sleep stages, EEG arousals, and respiratory episodes. These evaluations were conducted using the same CNN-LSTM architecture configuration ($s$ + $a$ + $r$) and full ($D = 8$) input montage (2$\times$EEG, EOG, EMG, oxygen saturation, airflow, abdominal and thoracic movements).
As it can be observed, using the three-component loss function results in a slight improvement in performance, with a $\kappa$ of 0.80, compared to a $\kappa$ of 0.78 achieved with the MSE-only loss. In addition, the training time is halved when utilizing the combined loss function, with the number of training epochs reducing from 51 to 31.

Tables \ref{local_performance} and \ref{external_performance} present the local and external performance metrics of the ten scheduled experiments, respectively. Each experiment uses several input combinations ($D$) and target output assemblies ($E$) to assess the resulting impact on the performance. 
From Table \ref{local_performance} it can be seen that the best average performance is achieved for the sleep staging task, with $\kappa$ of 0.82. Notably, there are no significant changes in performance for this specific task when varying the number of input channels or modifying the combination of sleep events predicted at the output ($\kappa$ ranges 0.79-0.83). 
In contrast, a positive trend is observed in the detection of respiratory episodes. Taking as reference the ($s$ + $r$) assembly, we observe enhanced performance as additional channels are incorporated to the input montage (from $\kappa$ = 0.48 when $D$ = 4, up to $\kappa$ = 0.65 when $D$ = 8). However, enlarging the number of input channels appears to have a detrimental effect on the detection of EEG arousals. That is, taking ($s$ + $a$) as reference, we see $\kappa$ going from 0.70 ($D$ = 4) down to 0.47 ($D$ = 8).

Figures \ref{fig:local_mae} and \ref{fig:external_mae} illustrate the impact of increasing the number of input channels ($D$) on MAE when simultaneously predicting sleep stages, EEG arousals, and respiratory events, respectively for the local and external testing scenarios. In this case MAE refers to the classification MAE and the location MAE jointly. 
From Figure \ref{fig:local_mae}, it can be noticed that MAE for respiratory events diminishes with the inclusion of additional channels. However, the prediction of EEG arousals exhibits an inverse relationship, with higher MAE as $D$ increases, while the MAE for sleep stages remains somewhat constant. This follows a similar trend with $\kappa$ scores, however attending to the global MAE, depicted in yellow, we see now that adding extra channels to the input montage contributes positively to the global model's performance. This trend, however, disappears when evaluating the MAE in predicting the HMC-ISA external dataset. In this case we see certain improvement when going from $D$=4 to $D$=6, but observe a clear deterioration in performance when $D$ = 8, with perhaps the only exception for the detection of EEG arousals.

In general, when comparing global average scores between Tables \ref{local_performance} and \ref{external_performance} we observe that, while promising performance is achieved for the testing source dataset ($\kappa$ values of 0.82 for sleep stages, 0.61 for EEG arousals, and 0.58 for respiratory events) a significant drop is noticeable when evaluating in the external dataset scenario, as evidenced by reduced averaged kappa values of 0.45 for sleep stages, 0.23 for EEG arousals, and 0.21 for respiratory events.

\section{Discussion}
\label{discussion}

\begin{figure}[t]
\centering
\includegraphics[width=0.65\textwidth]{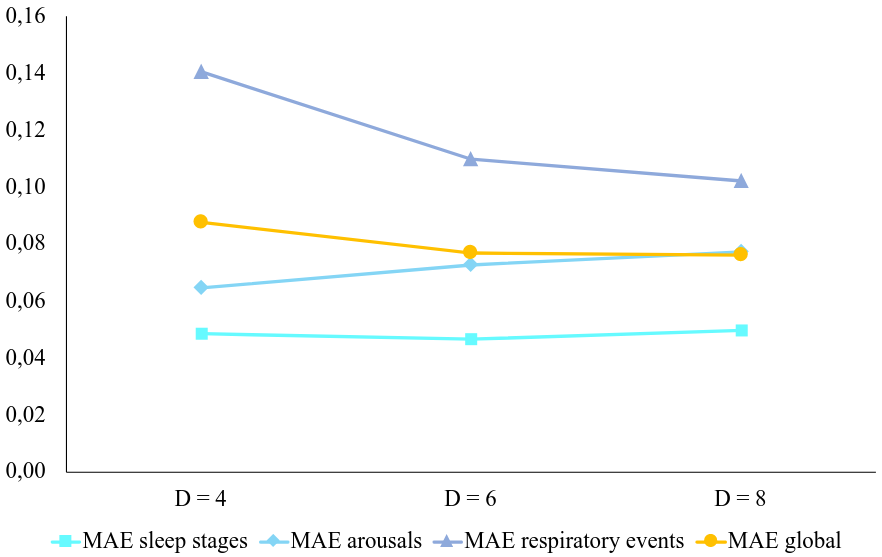}
\caption{Testing performance in the source database (SHHS) in terms of MAE. For each input channels $D$ combinations, global and event-specific MAEs are illustrated. In this case, MAE metric refers to the combination of the detection and classification tasks.}
\label{fig:local_mae}
\end{figure}

\begin{figure}[t]
\centering
\includegraphics[width=0.65\textwidth]{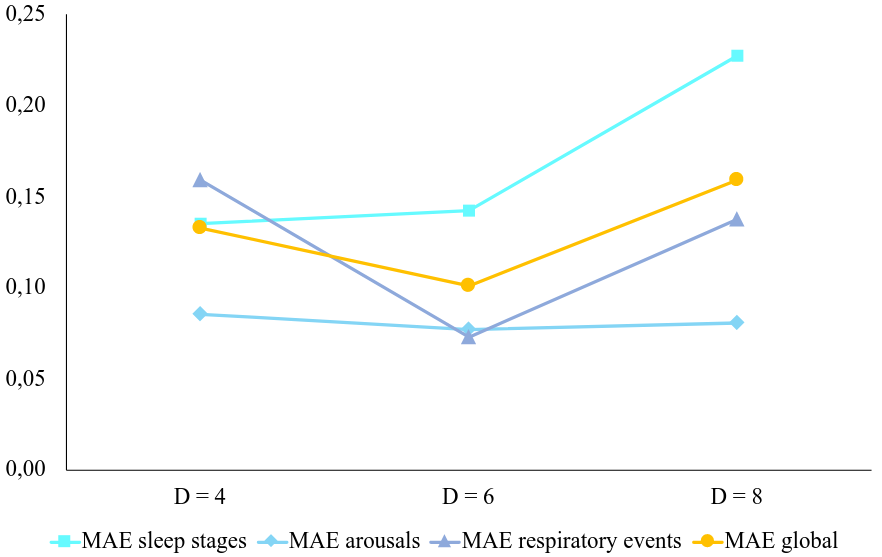}
\caption{Testing performance in the external database (HMC-ISA) in terms of MAE. For each combination of input channels $D$ global and event-specific MAEs are illustrated. In this case, MAE metric refers to the combination of the detection and classification tasks.}
\label{fig:external_mae}
\end{figure}

In this work we have proposed a novel multi-task deep-learning method for the jointly detection of sleep events and hypnogram construction, that carries out analysis in one single pass. More specifically, we have explored how our method performs when predicting different assemblies of sleeping events, and how the addition of different input signal montages affects the detection. 
For this purpose, we have considered different combinations of sleep stages, EEG arousals and respiratory episodes ($apneas$ and $hypopneas$) as target outputs. We have also explored the use of 4, 6 and 8 input signal combinations on predicting each task.

With regard to the expansion of the input dimensionality, our results have evidenced a significant improvement in the prediction of respiratory events, as outlined in Tables \ref{local_performance} and \ref{external_performance}. This aligns with expectations, since the added signal channels (blood oxygen saturation, airflow, and thoraco-abdominal movements) contain specific information related to the patient's breathing function. Notably, integrating oxygen saturation and airflow into the baseline 4-channel setup (2$\times$EEG, EOG, EMG) significantly increases Cohen's kappa scores for this specific task (from $\kappa=0.47$ to $\kappa=0.61$). However, the addition of thoracic-abdominal movement data, although positive, does not seem to have a drastic impact on the performance ($\kappa=0.61$ to $\kappa=0.65$). This suggests potential redundancy in the information provided by these channels in relation to the contribution of airflow and oxygen saturation. This observation motivates future research that could explore alternative configurations, such as using thoraco-abdominal movements instead of saturation and airflow, to assess their relative impact on the model performance.
In contrast, we have observed that EEG arousal detection performance does not benefit from the added extra channels. This effect may be attributable to the limited capacity or lack of degrees of freedom from the learning model, which needs to split resources across the multiple sub-tasks, potentially limiting the effectiveness in some of them, while others improve. Importantly, however, from a global perspective we have seem a positive performance effect by adding further structure to the output assembly.
Considering 4 input signals, for instance, our method has evidenced that performing multiple tasks simultaneously enhances sleep stages prediction in both local and external datasets (see Tables \ref{local_performance} and \ref{external_performance}). We have also observed a net improvement in the global MAE values for the full assembly configuration. These outcomes are therefore consistent with the principles of multi-task learning, which leverages the synergistic knowledge acquired from the related sub-tasks to improve overall model performance \cite{zhang2021survey}. 
Note that this is especially advantageous in this scenario, where task-specific data are limited, as the completion of one specific task assists the performance of others by encouraging the extraction of relevant and discriminatory features. 

The use of a combined loss function has also shown notable improvements in both performance and training speed, as compared to a standard single loss function approach (refer to Table \ref{table_loss}). This enhancement is attributable to the task-specific nature of each of the loss components, which facilitates more precise error minimization for each task, thereby guiding the model towards a more accurate and robust solution.
On the other hand, a relevant question that still deserves more attention regards the relative event-specific contribution to the loss function, in particular due to uneven class distribution and distinct characteristics of the different sleep event types. For example, as we have already discussed, we have seen in Figure \ref{fig:local_mae}, that MAE for scoring of EEG arousals increases inversely proportional to that of respiratory events when several tasks are considered for simultaneous detection. This could also suggest that our training database contains fewer examples of EEG arousals in comparison to the proportion of respiratory episodes and, therefore, our model tends to specialize in the detection of the latter.

Moreover, significant inter-database variability is expected in relation to population-specific phenotype, scoring methods, human subjectivity, or specific montage configurations \cite{alvarez-estevez_inter-database_2021}. Hence, bias might likewise be induced in automatic learning models towards the most frequently represented examples and/or event characteristics of a certain database.
In fact, our experimentation has revealed significant degradation in the prediction performance when the model targets an external database (Figure \ref{fig:external_mae}). Here, we should note that an increased focus towards respiratory inputs during training (i.e. more specialization in these events) correlates with higher MAE in the external validation, which suggests that source and target databases could contain a different distribution of this type of events. Similar downgrading effects have been previously documented, mostly related to sleep staging tasks throughout the literature \cite{biswal2018expert, Cesari2021, alvarez-estevez_challenges_2023, olesen_automatic_2021}. In this regard, the straightforward solution to enhance model generalization involves increasing data amount and heterogeneity by integrating multiple databases in the training dataset. However, such data-centralization approach may be technically challenging, cost-expensive, difficult to scale over time, and could compromise sensitive information. As an alternative, decentralized learning emerges as an interesting option \cite{anido2023decentralized}.

In future work, we aim to extend our multi-task learning approach to further databases and explore different decentralized deep-learning methods for enhancing inter-database generalization. Moreover, we would like to experiment with alternative neural network architectures, hyper-parameterization, and more specialized loss functions, that could help mitigate some of the possibly induced learning biases, and tackle potential model capacity constraints.
We would also like to expand detection capabilities and intend to include the simultaneous scoring of additional sleep-related events, such as limb movements. Finally, we still need to work on global fine-tuning and carry out further experimentation with the use of different window sizes and sub-window configurations to improve the overall flexibility of our method.

\section{Conclusions}
\label{conclusions}
In this work we have presented a first approximation towards a flexible and adaptive method for joint detection of sleep events and hypnogram construction in a single pass. 
We have confirmed that our method supports different input configurations, thus adapting to different PSG montages. Moreover, our output configuration is flexible and scalable, able to accommodate several multi-task targets. Finally, it integrates all the available PSG information to perform the analysis and annotate several sleep events at once, facilitating clinicians' practice.

\section{Acknowledgments}
This study has been supported by project \mbox{RYC2022-038121-I}, funded by \mbox{MCIN/AEI/10.13039/501100011033} and European Social Fund Plus (ESF+) and project \mbox{PID2023-147422OB-I00} funded by \mbox{MCIU/AEI/10.13039/501100011033} and by the European FEDER program. CITIC, as a center accredited for excellence within the Galician University System and a member of the CIGUS Network, receives subsidies from the Department of Education, Science, Universities, and Vocational Training of the Xunta de Galicia. Additionally, it is co-financed by the EU through the FEDER Galicia 2021-27 operational program (Ref. ED431G 2023/01).

\bibliographystyle{ieeetr}
\bibliography{bib}

\end{document}